\def\aap{\mbox{\bf{Astron.\ Astrophys.}}}
\def\apj{\mbox{\bf{Astrophys.\ J.}}}
\def\apjl{\mbox{\bf{Astrophys.\ J.\ L.}}}
\def\aj{\mbox{\bf{Astron.\ J.}}}

\def\pasj{\mbox{\bf{Pub.\ Astronom.\ Soc.\ Japan}}}
\def\pasp{\mbox{\bf{Pub.\ Astron.\ Soc.\ Pacific}}}
\def\mnras{\mbox{\bf{Mon.\ Not.\ Roy.\ Astron.\ Soc.}}}

\newcommand{\msun}{\mbox{${\rm M}_{\odot}$}}
\newcommand{\Msun}{\mbox{${\rm M}_{\odot}$}}

\newcommand{\tdis}{\mbox{$t_{\rm dis}$}}

\newcommand{\tenc}{\mbox{$t_{\rm enc}$}}

\newcommand{\ec}{\mbox{$E_{\rm 0}$}}

\newcommand{\Mcl}{\mbox{$M_{\rm c}$}}
\newcommand{\Mc}{\mbox{$M_{\rm c}$}}
\newcommand{\Mdot}{\mbox{$\dot{M}_{\rm c}$}}
\newcommand{\Mn}{\mbox{$M_{\rm n}$}}
\newcommand{\Mcloud}{\mbox{$M_{\rm n}$}}

\newcommand{\kms}{\mbox{km$\,$s$^{-1}$}}
\newcommand{\pcmsq}{\mbox{${\rm pc}^{-2}$}}

\newcommand{\rhon}{\mbox{$\rho_{\rm n}$}}
\newcommand{\rh}{\mbox{$r_{\rm h}$}}
\newcommand{\rt}{\mbox{$r_{\rm t}$}}
\newcommand{\rc}{\mbox{$r_{\rm c}$}}

\newcommand{\reff}{\mbox{$r_{\rm eff}$}}
\newcommand{\rv}{\mbox{$r_{\rm v}$}}
\newcommand{\rrms}{\mbox{$\bar{r^2}$}}
\newcommand{\rhs}{\mbox{$r^2_{\rm h}$}}
\newcommand{\acsq}{\mbox{${a^2_{\rm c}}$}}

\newcommand{\vrms}{\mbox{$\sigma_{\rm rms}$}}
\newcommand{\vcn}{\mbox{$\sigma_{\rm cn}$}}
\newcommand{\Rn}{\mbox{$R_{\rm n}$}}
\newcommand{\sigman}{\mbox{$\Sigma_{\rm n}$}}
\newcommand{\sigmamol}{\mbox{$\Sigma_{\rm mol}$}}

\newcommand{\vmax}{\mbox{$V_{\rm max}$}}
\newcommand{\dr}{\mbox{${\rm d}$}}

\newcommand{\ac}{\mbox{$a_{\rm c}$}}
\newcommand{\an}{\mbox{$a_{\rm n}$}}

\newcommand{\ncloud}{\mbox{$n_{\rm clouds}$}}
\newcommand{\f}{\mbox{$0.22$}}
\newcommand{\de}{\mbox{$\Delta E/|E_0|$}}
\newcommand{\dm}{\mbox{$\Delta M/M_0$}}
\newcommand{\dn}{\mbox{$\Delta N/N_0$}}

\newcommand{\tdissingle}{\mbox{$t_{\rm dis}^{\rm single}$}}
\newcommand{\pcrit}{\mbox{$p_{\rm crit}$}}

\def\aj{AJ}%
\def\apj{ApJ}%
\def\apjl{ApJ}%
\def\aap{A\&A}%
\def\mnras{MNRAS}%
\def\pasp{PASP}%
\def\pasj{PASJ}%
\documentclass[useAMS,usenatbib]{mn2e}
\usepackage{latexsym,graphicx,natbib,amssymb}

\title[Star clusters disruption by giant molecular clouds]
{Star cluster disruption by giant molecular clouds}

\author[M. Gieles et al.]
  {M.~Gieles$^{1,2}$, S.~F.~Portegies Zwart$^2$, H. Baumgardt$^3$, E.~Athanassoula$^4$,
  \newauthor  H.~J.~G.~L.~M. Lamers$^{1,5}$, M. Sipior$^2$,  and J. Leenaarts$^1$\\
  $^1$ Astronomical Institute, Utrecht University, 
  Princetonplein 5, 3584 CC Utrecht, The Netherlands \\
  $^2$ Astronomical Institute `Anton Pannekoek', University of Amsterdam, 
  Kruislaan 403, 1098 SJ Amsterdam, The Netherlands, \\\
  Section Computational Science, 
  University of Amsterdam, Kruislaan 403, 1098 SJ, The Netherlands \\
  $^3$ Sternwarte, University of Bonn, Auf dem H\"{u}gel 71, 53121 Bonn, Germany\\
  $^4$ Observatoire de Marseille, 2 Place le Verrier, 
  13248 Marseille Cedex 4, France\\
  $^5$ SRON Laboratory for Space Research, Sorbonnelaan 2, NL-3584 CA Utrecht, The Netherlands
}
\date{Released 2006 Xxxxx XX}

\pagerange{\pageref{firstpage}--\pageref{lastpage}} \pubyear{2002}

\def\LaTeX{L\kern-.36em\raise.3ex\hbox{a}\kern-.15em
    T\kern-.1667em\lower.7ex\hbox{E}\kern-.125emX}

\begin{document}         
\maketitle

   \begin{abstract} We investigate encounters between giant molecular
      clouds (GMCs) and star clusters. We propose a single expression
      for the energy gain of a cluster due to an encounter with a GMC,
      valid for all encounter distances and GMC properties. This
      relation is verified with $N$-body simulations of cluster-GMC
      encounters, where the GMC is represented by a moving analytical
      potential. Excellent agreement is found between the simulations
      and the analytical work for fractional energy gains of $\Delta
      E/|E_0|<10$, where $|E_0|$ is the initial total cluster
      energy. The fractional mass loss from the cluster scales with
      the fractional energy gain as $(\Delta M/M_0)=f(\Delta
      E/|E_0|)$, where $f\simeq0.25$.  This is because a fraction
      $1-f$ of the injected energy goes to the velocities of escaping
      stars, that are higher than the escape velocity. We therefore
      suggest that the disruption time of clusters, $\tdis$, is best
      defined as the time needed to bring the cluster mass to zero,
      instead of the time needed to inject the initial cluster
      energy. We derive an expression for $\tdis$ based on the mass
      loss from the simulations, taking into account the effect of
      gravitational focusing by the GMC. Assuming spatially
      homogeneous distributions of clusters and GMCs with a relative
      velocity dispersion of \vcn, we find that clusters loose most of
      their mass in relatively close encounters with high relative
      velocities ($\sim2\,\vcn$). The disruption time depends on the
      cluster mass ($\Mc$) and  half-mass radius ($\rh$) as
      $\tdis~=~2.0\,S\,\left(\Mc/10^4\,\msun\right)\,\left(3.75\,\mbox{pc}/\rh\right)^3{\mbox{Gyr}}$,
      with $S\equiv1$ for the solar neighbourhood and $S$ scales with
      the surface density of individual GMCs ($\sigman$) and the
      global GMC density ($\rhon$) as
      $S\propto(\sigman\rhon)^{-1}$.  Combined with the observed
      relation between \rh\ and \Mc, i.e. $\rh \propto
      \Mcl^{\lambda}$, \tdis\ depends on \Mc\ as
      $\tdis\propto\Mc^\gamma$. The index $\gamma$ is then defined as
      $\gamma=1-3\lambda$.  The observed shallow relation between cluster
      radius and mass (e.g. $\lambda\simeq0.1$), makes the value of
      the index $\gamma=0.7$ similar to that found from observations
      and from simulations of clusters dissolving in tidal fields
      ($\gamma\simeq0.62$).  The constant of 2.0 Gyr, which is the
      disruption time of a $10^4\,\msun$ cluster in the solar
      neighbourhood, is about a factor of 3.5 shorter than found from
      earlier simulations of clusters dissolving under the combined
      effect of galactic tidal field and stellar evolution. It is
      somewhat higher than the observationally determined value of 1.3
      Gyr. It suggests, however, that the combined effect of tidal
      field and encounters with GMCs can explain the lack of old open
      clusters in the solar neighbourhood. GMC encounters can also
      explain the (very) short disruption time that was observed for
      star clusters in the central region of M51, since there $\rhon$
      is an order of magnitude higher than in the solar neighbourhood.
\end{abstract}

%%%%%%%%%%%%%%%%%%%%%%%%%%%%%%%%%%%%%%%%%%%%%%%%%%%%%%%%%%%%%%%%%%%
\section{Introduction}
\label{sec:introduction}

Star clusters are subjected to various disruptive effects, which
prevent low mass star clusters ($\Mcl<\,10^4\,\msun$) from surviving
for a Hubble time. From observations this was first noted by
\citet{1958RA......5..507O} and later by \citet{1971A&A....13..309W},
who showed that there is a lack of open star clusters older than a few
Gyr in the solar neighbourhood. Later, more quantitative results were
obtained for the life time of star clusters
(e.g. \citealt{1988PASP..100..576H} and \citealt{2006MNRAS.366..295D}
for clusters in the LMC; \citealt{2003MNRAS.338..717B} for the SMC and
M33;
\citealt{2005A&A...441..117L} and \citealt{2006A&A...445..545P} for the
solar neighbourhood and \citealt{2005A&A...441..949G} for the central
region of M51). From the theoretical side, the evolution and
disruption of star clusters has been the subject of many studies using
a variety of techniques to solve the gravitational $N$-body
problem (e.g. \citealt*{1972ApJ...176L..51O, 1990ApJ...351..121C,
1997ApJ...474..223G, 2000ApJ...535..759T, 2003MNRAS.340..227B}).

\citet*{2005A&A...429..173L} compared observationally determined 
disruption times in four galaxies to the corresponding times following
from the $N$-body simulations of clusters in tidal fields of
\citet{2003MNRAS.340..227B} and
\citet{1998A&A...337..363P,2002ApJ...565..265P}. They found that,
based on the results of the simulations, the disruption time ($\tdis$)
should depend on the cluster mass $\Mc$ as $\tdis \propto
[\Mcl/\ln(\Mcl)]^{0.75} \propto \Mcl^{0.62}$. The index (0.62), which
is referred to as $\gamma$, was also determined from the observed age
and mass distributions of different cluster populations
\citep{2003MNRAS.338..717B} and the mean value from the observations
is $\bar{\gamma}=0.62\pm0.06$. The value of $\tdis$ scales with the
disruption time of a $10^4\,\msun$ cluster as
$\tdis=t_4\,(\Mcl/10^4\,\msun)^\gamma$. Good agreement for $t_4$
between simulations and observations was found only for the star
clusters in the SMC. In the solar neighbourhood, the observed $t_4$
for open clusters ($t_4=1.3\,$Gyr, see \citealt{2005A&A...441..117L})
is a factor of 5 lower than expected from the disruption time due to
the galactic tidal field, combined with a realistic stellar mass function and
with stellar evolution ($t_4 = 6.9\,$Gyr, see
\citealt{2003MNRAS.340..227B}). An even larger disagreement of almost
an order of magnitude was found for $\tdis$ of clusters in the central
region of M51
\citep{2005A&A...441..949G}.

The fact that the observed disruption times are shorter could be a
result of time-dependent external perturbations, such as spiral arm
passages and encounters with giant molecular clouds (GMCs), which were
ignored in the simulations of
\citet{2003MNRAS.340..227B}. \citet{1987MNRAS.224..193T} and
\citet{1991MmSAI..62..909T} showed that a single encounter with a
massive GMC could disrupt a star cluster of a few hundred $\Msun$ and
\citet*{gieles06a} showed that in the solar neighbourhood passages of
spiral arms contribute significantly to the disruption of open
clusters.

In this study we investigate the disruption of clusters by encounters
with GMCs. We want to quantify whether GMCs can explain the difference
between the observed values of $t_4$ and the ones following from the
simulations without the time-dependent external effects. Moreover,
would simulations including encounters with GMCs preserve the
numerical value of the index $\gamma\simeq0.62$, where the
observations and previous simulations already agree?

The paper is organised as follows: In \S~\ref{sec:initial} we
introduce the initial conditions for the clusters and GMCs and define
the parameters involved in a cluster-GMC encounter. In
\S~\ref{sec:analytical} we give analytical calculations of the
energy gain for a cluster due to various types of encounters. These
analytical formulae for the energy gain are verified with $N$-body
simulations and compared to the mass loss of the cluster in
\S~\ref{sec:nbody}. The results of the $N$-body simulations are used
in \S~\ref{sec:disruption} to derive an expression for the cluster
disruption time in environments with different GMC densities. The conclusions
and discussion are presented in \S~\ref{sec:conclusion}.

%%%%%%%%%%%%%%%%%%%%%%%%%%%%%%%%%%%%%%%%%%%%%%%%%%%%%%%%%%%%%%%%%%%
\section{Initial conditions}
\label{sec:initial}

%_________________________________________________________________
\subsection{GMC properties}
In this study we consider encounters between molecular clouds and star
clusters. Since 90\% of the total molecular gas mass in our Galaxy is
in giant molecular clouds (GMCs) with mass $\Mn>10^4\,\msun$
\citep{1987ApJ...319..730S} and encounters with low mass clouds do not
affect a star cluster much (e.g. \citealt{1985IAUS..113..449W}), we
consider only the massive GMCs. In \S~\ref{sec:analytical} we show
that the cluster energy gain due to an encounter with a GMC scales
with the GMC mass squared ($\Mn^2$), supporting our assumption to
consider only the massive clouds (i.e. the GMCs).
\citet{1987ApJ...319..730S} showed that there is a relation between
the size and mass of galactic GMCs of

\begin{equation}
\Mn = 540\,\left(\frac{\Rn}{{\rm pc}}\right)^2\,\,\msun,
\label{eq:mrclouds}
\end{equation}
where $\Mn$ is the mass of the GMC and $\Rn$ is its
radius\footnote{Throughout this paper we use the subscript ``$n$''
(from nebula  or nuage) to indicate the parameters of the GMC and ``$c$'' for
the cluster.}. The internal density profile is of the form
$\rho(r)\propto r^{-1}$. Eq.~\ref{eq:mrclouds} implies a constant mean
surface density of $\sigman = 170\,\msun\pcmsq$.  

\begin{figure}
    \includegraphics[width=8cm]{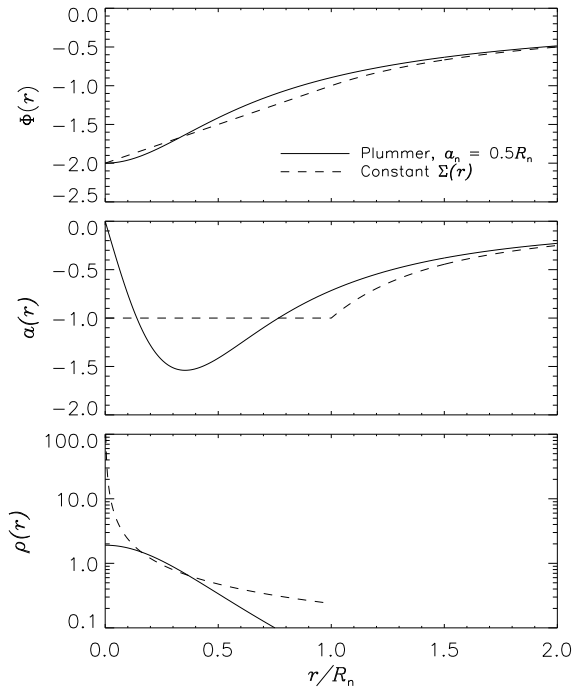}
    \caption{Comparison between the radial profiles of the potential
    ({\it top}), acceleration ({\it middle}) and density ({\it
    bottom}) for a constant surface density GMC with $\Mn=\Rn = 1$, 
    following from Eq.~\ref{eq:mrclouds} and $\rhon \propto r^{-1}$
    (dashed line) and for a Plummer model with radius $\an =
    0.5\,\Rn$ (full line).}

    \label{fig:pad}
\end{figure}

We use the spherically symmetric Plummer model
\citep{1911MNRAS..71..460P} to describe the GMC, since it is 
mathematically convenient to use in analytical calculations
(\S~\ref{sec:analytical}). The potential of this model is
\begin{equation}
\Phi(r)=-G\,M/\sqrt{r^2+a^2},
\label{eq:plummer}
\end{equation}
where $G$ is the gravitational constant, $M$ is the total mass and $a$
is the Plummer radius. If we choose the Plummer radius of a GMC equal
to half the radius from Eq.~\ref{eq:mrclouds}, e.g. $\an = 0.5\,\Rn$,
the resulting profile is very similar to what would follow from a
density profile of the form $\rho(r)\propto r^{-1}$. In
Fig.~\ref{fig:pad} we compare the descriptions for a Plummer model
(Eq.~\ref{eq:plummer}) and constant surface density profile
(Eq.~\ref{eq:mrclouds}) for the potential in the top panel, the
acceleration in the middle panel and the density profile in the bottom
panel. The Plummer model has the advantage that the first derivative
of the force is differentiable for each $r$, which is necessary for
the $N$-body simulations.

%_________________________________________________________________
\subsection{Star cluster properties}
\label{subsec:cluster}
For the star cluster we also choose a Plummer model
(Eq.~\ref{eq:plummer}). The number of particles is $N=2048$ with a
total mass of $\Mcl=1000\,\Msun$ and all stars the same mass,
resulting in a realistic mean stellar mass of 0.49\,\msun\
(e.g. \citealt{2001MNRAS.322..231K}). The virial radius is set to
2~pc. For a Plummer model, the virial radius relates to the Plummer
radius of the cluster as $\rv = 16\,\ac/(3\pi)$. The cluster half-mass
radius relates to $\ac$ as $\rh
\simeq 1.305\,\ac$. Therefore, the values of $\ac$ and $\rh$ in
physical units are 1.18 pc and 1.54 pc, respectively.  The energy
gain of a cluster due to an encounter with a GMC will be compared to
the total initial cluster energy, e.g. the sum of kinetic and
potential energy. This is defined as $E_0=-\eta\,G\Mc^2/(2\rh)$, with
$\eta\simeq0.4$, depending on the cluster model. For a Plummer model
$E_0=(3\pi/64)\,G\Mc^2/\ac$, hence $\eta\simeq0.38$.

%_________________________________________________________________
\subsection{Encounter definitions}
\label{subsec:relations}
\begin{figure}
    \includegraphics[width=8cm]{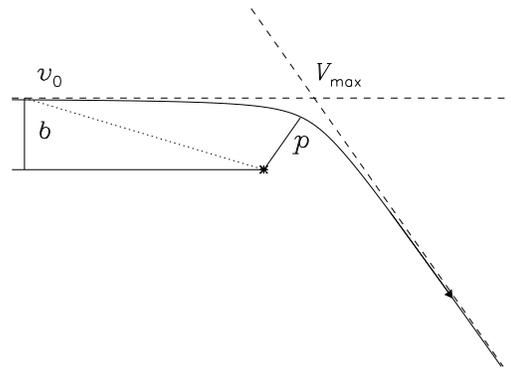}

    \caption{Motion of the reduced particle in a hyperbolic
    encounter. The impact parameter $b$ and the velocity at infinity $v_0$ are denoted on the
    left side of the picture and the encounter parameter $p$ and
    maximum relative velocity $\vmax$ are indicated at the point
    of closest approach.}

\label{fig:impact_encounter}

\end{figure}

Two particles with a relative velocity at infinity that is non-zero
will experience a hyperbolic encounter. %A fictitious particle, called
%the reduced particle with mass $\mu \equiv mM/(M+m)$, then moves in a
%static potential with mass $M+m$. 
The motion of the reduced mass particle is schematically represented
in Fig.~\ref{fig:impact_encounter}. The impact parameter and the velocity
at infinity are referred to as $b$ and $v_0$, respectively. The
distance of closest approach and the maximum velocity are called the
encounter parameter $p$ and $\vmax$, respectively.  Conservation of
angular momentum relates $v_0$ and $\vmax$ as

\begin{equation}
\vmax=v_0\frac{b}{p}.
\label{eq:v}
\end{equation}
Consider a GMC with a Plummer model according to
Eq.~\ref{eq:plummer} and mass $\Mn$ and a cluster with mass $\Mc$
which is a point source. From conservation of energy and
Eq.~\ref{eq:v}, it follows that $b$ and $p$ are related as
\begin{equation}
b = p\,\sqrt{1+\frac{2G(\Mn+\Mc)}{v_0^2\sqrt{p^2+\an^2}}}.
\label{eq:pb}
\end{equation}
Note that $b$ is now a mixed function of encounter and impact
variables. In the next sections we will show that the encounter
parameter $p$ and $\vmax$ determine the internal energy gain. Since we want to
perform $N$-body simulations that result in an encounter with a
certain desired $p$ and $\vmax$, we use Eq.~\ref{eq:v} and
Eq.~\ref{eq:pb} to express $p$ and $\vmax$ in terms of $b$ and $v_0$. In
addition, in our simulations we will not start with the GMC at
infinity, hence we will need to take into account the potential energy
contribution at the beginning of the simulation. Since the final
expressions for $b$ and $v_0$ as a function of $p, \vmax, \Mn$ and
$\Mcl$ are quite complex, but follow from simple mathematical
exercises, we do not give them here.

There exists an analytical expression for the energy gain of a cluster
due to an head-on encounter and due to a distant encounters. In the
next section these relations will be extended to the full range of
encounter parameters, i.e. including close encounters. The exact boundary
between close and distant encounters is hard to define. Here we will
distinguish three types of encounters: (1) head-on encounters for
which $p=0$; (2) close encounters, where $0<p/\an<5$ and (3) distant
encounters, which is for $p\ge5\an$.

%%%%%%%%%%%%%%%%%%%%%%%%%%%%%%%%%%%%%%%%%%%%%%%%%%%%%%%%%%%%%%%%%%%
\section{Cluster energy gain and mass loss due to an encounter with a GMC}
\label{sec:analytical}

%_________________________________________________________________
\subsection{Distant encounters}
\label{subsec:distant}
Spitzer (\citeyear{1958ApJ...127...17S}, hereafter S58) derived an expression for the energy gain 
of a cluster due to a distant encounter with a GMC.  He assumed that
the GMC is a point mass and that the encounter velocity is much
faster than the internal velocity of the stars in the cluster, which
is referred to as the {\it impulsive approximation}. The cluster
energy gain can then be expressed in the encounter, GMC and cluster
parameters as

\begin{equation}
\Delta E = \frac{1}{3}\,\left(\frac{2G \Mn}{p^2 \vmax}\right)^2\,\Mc\rrms.
\label{eq:despitzer}
\end{equation}
Here $\rrms$ is the mean-square position of the stars in the
cluster\footnote{S58 makes no distinction between $b,v_0$ and
$p,\vmax$. In \S~\ref{sec:nbody} we show that $p$ and
$\vmax$ are the most convenient ones to use for calculating the energy gain.}.

\citet{1991MmSAI..62..909T} used combined  
hydro-dynamical and $N$-body calculations of a GMC-cluster encounter,
to show that Eq.~\ref{eq:despitzer} is very accurate. He also notes
the inconvenience of the ill defined \rrms\ parameter. In principle
$\rrms$ follows from the cluster profile, but this variable is very
sensitive to variations of the positions of stars far from the cluster
centre, hence it is not constant during the simulation because of
scattering of stars out of the cluster. Even in simulations covering a short time span,
$\rrms$ will increase due to stars that get at larger distances from
the centre due to 2-body encounters in the
core. 

\citet{1985ApJ...295..374A} have shown with $N$-body
calculations of encounters between two Plummer models with radii $a_1$
and $a_2$, that the predictions of Eq.~\ref{eq:despitzer} hold when
$p\ge\mbox 5\,{\rm max}(a_{1}, a_{2})$. In \S~\ref{sec:nbody} we
examine with $N$-body simulations in what regime the results of
Eq.~\ref{eq:despitzer} are valid, and at what moment during the
encounter the value for \rrms\ should be taken.

%_________________________________________________________________
\subsection{Head-on encounters}
\label{subsec:headon}

Gravitational focusing will deflect many cluster orbits which start
with $b>5\,\an$ to encounters with $p<5\,\an$, for which the relation
of S58 (Eq.~\ref{eq:despitzer}) does not hold. To obtain an expression
for the energy gain for close encounters, we first explore the
opposite of distant encounters, namely head-on encounters
(i.e. $p=0$). In those encounters the extent of the GMC has to be
taken into account, since assuming a point mass GMC would severely
overestimate the energy gain. The energy gain in close encounters
should be an interpolation between the distant encounter and
head-on results.

Following \citet{1987gady.book.....B} (Chap. 7.2, hereafter BT87) we
can derive the cluster energy gain for an head-on encounter between
a cluster and a GMC, both described by a Plummer model. Let $(R,z)$ be
cylindrical coordinates, with the $z$-axis coinciding with the GMC
trajectory. Due to symmetry, the stars in the cluster get a
velocity increase only perpendicular to the trajectory of the GMC, of size
\begin{equation}
\Delta v\,(R) = -\frac{2 G \Mcloud R}{\vmax(R^2+\an^2)},
\label{eq:dv}
\end{equation}
where analogous to the case of the distant encounters we use $\vmax$ for the
relative velocity between the cluster and the GMC. Eq.~\ref{eq:dv} can be
used to calculate the energy gain of the cluster:
\begin{equation}
\Delta E = \pi\,\int_0^\infty [\Delta v\,(R)]^2\,\Sigma(R)\,R\,\dr R.
\end{equation}
Here $\Sigma(R)$ is the surface density of the cluster obtained by
projecting parallel to the line of motion of the GMC. Since the cluster
is described by a Plummer model, $\Sigma(R)$ is

\begin{equation}
\Sigma(R) = \frac{\Mc \ac^2}{\pi}\,(R^2 + \ac^2)^{-2},
\end{equation}
so the energy gain of the cluster due to an encounter between two Plummer models is

\begin{equation}
\Delta E = \left(\frac{2 G \Mcloud \ac}{\vmax}\right)^2\,M_c\int_0^\infty \frac{R^3}{(R^2+\an^2)^2\,(R^2+\ac^2)^2}\dr R.
\label{eq:de1}
\end{equation}
The result of the integration is a function of $\ac$ and $\an$ and is
equal to $1/12a^4$ when $\an = \ac = a$. In that case, Eq.~\ref{eq:de1}
reduces to the result found by BT87:

\begin{equation}
\Delta E_{\rm BT} = \frac{1}{3}\left(\frac{G\Mcloud}{\vmax\,a}\right)^2\,M_c.
\label{eq:debt}
\end{equation}
When $\an \ne \ac$, the result of the integration is a function of
$\ac$ and $\an$. Introducing a new variable $\chi\equiv\an/\ac$, the
ratio of the GMC and cluster radii, Eq.~\ref{eq:de1} can be written as
\begin{equation}
\Delta E = \frac{1}{3}\left(\frac{G \Mcloud}{\vmax\,\ac}\right)^2 M_c\frac{12(\chi^2-1)-6(\chi^2+1)\,\ln(\chi^2)}{(1-\chi^2)^3},
\end{equation}
where in the first factor we recognise the result for $\ac = \an$
(e.g. Eq.~\ref{eq:debt}) and the rest is a correction factor $C(\chi)$
which is only a function of $\chi$. We can write a more general
expression for the energy gain of head-on encounters

\begin{equation}
\Delta E = \Delta E_{\rm BT}\,\,C(\chi), 
\label{eq:deheadon}
\end{equation}
with
\begin{equation}
C(\chi) = \left\{ \begin{array}{ll} 
\frac{12(\chi^2-1)-6(\chi^2+1)\,\ln(\chi^2)}{(1-\chi^2)^3} & \textrm{for $\chi\ne1$}\\
1                                              & \textrm{for $\chi=1$}
\end{array}\right.
\label{eq:cx}
\end{equation}
The expression for $C(\chi)$ is rather complicated, but for realistic
values of $\chi$ (i.e. $2<\chi<10$), $C(\chi)$ can be approximated
with high accuracy by $C(\chi) \simeq 2\,\chi^{-3}$ (see Fig.~\ref{fig:head_on}).

With Eq.~\ref{eq:deheadon} and Eq.~\ref{eq:cx} we have a general
expression for the energy gain of head-on GMC-cluster encounters,
when both are described by Plummer models, valid for any radius and mass of
the cluster and the GMC. In \S~\ref{subsec:headon_nbody} we will
confront Eq.~\ref{eq:deheadon} and Eq.~\ref{eq:cx} with results from
$N$-body simulations.

%_________________________________________________________________
\subsection{A general expression for the cluster energy gain}
\label{subsec:detotal}

The energy gain for a close encounter, i.e. an encounter for which the
cluster moves through the cloud within a few $\an$ from the GMC
centre, should be a smooth interpolation between the result for a
head-on (Eq.~\ref{eq:deheadon}) and distant encounters
(Eq.~\ref{eq:despitzer}). When replacing $4\rrms/p^4$ in
Eq.~\ref{eq:despitzer} by $C(\chi)/\ac^2$, we get
the result for head-on encounter (Eq.~\ref{eq:deheadon}). One way of
connecting the two is to add a term in the denominator of
Eq.~\ref{eq:despitzer}, preventing the result to diverge to infinity at
$p=0$, so that it converges to Eq.~\ref{eq:deheadon}, is

\begin{equation}
\Delta E  = \frac{1}{3}\frac{4\rrms}{\left(p^2+\sqrt{4\,\rrms\acsq/C(\chi)}\right)^2}\,\left(\frac{G\Mcloud}{\vmax}\right)^2\,M_c.
\label{eq:detot}
\end{equation}
If we use the approximation given in \S~\ref{subsec:headon} for
$C(\chi)$ (i.e. $C(\chi)\simeq2\,\chi^{-3}$) and the fact that
$\sqrt{\rrms}\simeq1.8\,\rh$, $\ac\simeq\rh/1.305$ and the real GMC radius
$\Rn$ relates to $\an$ as $\Rn=2\,\an$, we can rewrite
Eq.~\ref{eq:detot} as 
%for plummer with cut off r_rms = 1.4, without r_rms = 1.77
\begin{equation}
\Delta E  \simeq \frac{4.4\,\rhs}{\left(p^2+\sqrt{\rh\,\Rn^{3}}\right)^2}\,\left(\frac{G\Mcloud}{\vmax}\right)^2\,M_c.
\label{eq:detotrh}
\end{equation}
In the next section, this relation will be compared to results of $N$-body simulations for
encounters with $p$ ranging from 0 to $10\,\an$.

%%%%%%%%%%%%%%%%%%%%%%%%%%%%%%%%%%%%%%%%%%%%%%%%%%%%%%%%%%%%%%%%%%%
\section{Validation with $N$-body simulations}
\label{sec:nbody}

%_________________________________________________________________
\subsection{Description of the code}
\label{subsec:code}
The $N$-body calculations were carried out by the $kira$
integrator, which is part of the Starlab software environment
(\citealt{1996ApJ...467..348M}; \citealt{2001MNRAS.321..199P}). $Kira$
uses a fourth-order Hermite scheme and includes special treatments of
close two-body and multiple encounters of arbitrary complexity. The
special purpose GRAPE-6 systems \citep{2003PASJ...55.1163M} of the
University of Amsterdam are used to accelerate the calculation of
gravitational forces between stars.

The potential, acceleration and jerk of the stars due to the GMCs are
derived from Eq.~\ref{eq:plummer} and calculated for each star
individually, based on its position and velocity with respect to the
GMC. We assume the GMCs are moving potentials which do not
interact. Indeed, we are interested in the star cluster and since we
consider GMCs that are $\sim 100-1000$ times more massive than the
cluster, the gravitational pull from the star cluster on the GMCs is
negligible and will therefore not affect its orbit. Although GMCs
have finite life times ($\simeq10-30\,$Myr), the encounter durations
are sufficiently small ($\simeq 1\,$Myr) to not be affected by this.

\begin{figure*}
    \includegraphics[width=17cm]{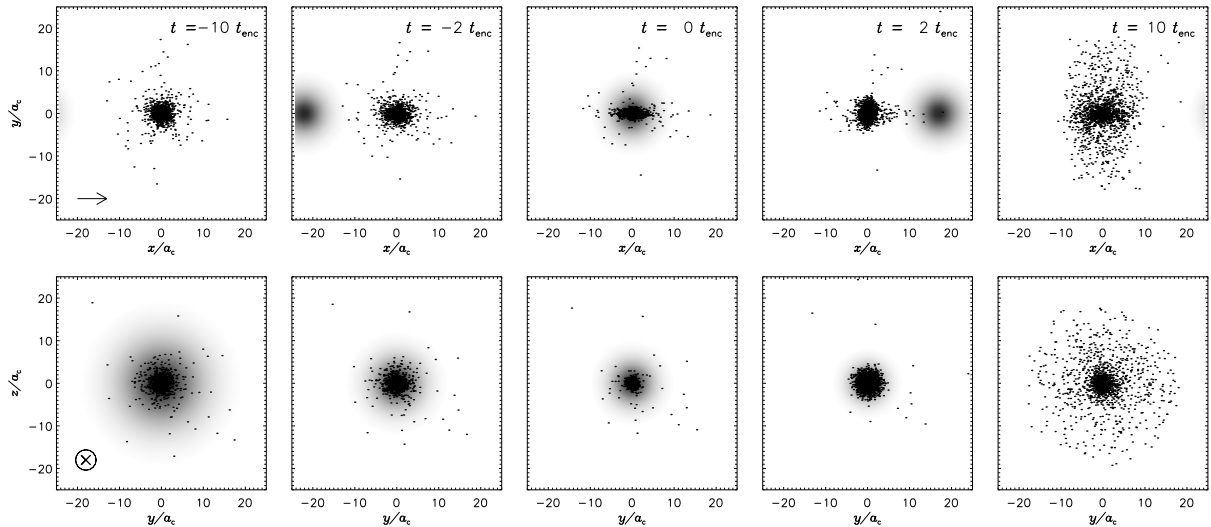}

    \caption{Snapshots of a star cluster undergoing a head-on
    encounter with a GMC of $\Mn=10^4\,\msun$ with $\vmax=20\,\vrms$.
    The motion of the GMC is along the $x$-axis the line of sight is
    perpendicular to ({\it top panels}) and along ({\it bottom
    panels}) the GMC trajectory. The arrows in the left lower corner
    of the left panels are parallel to the direction of motion of the
    GMC. The GMC is shown with grey shades based on the surface
    density of a GMC with $\an=5.8\,\ac$.  The time with respect to
    the moment of encounter is indicated in each panel of the top row,
    where $\tenc\equiv 2\an/\vmax$. The snapshots are taken in the
    centre-of-mass frame of the cluster and the viewing distance is
    $40\,\ac$.}

\label{fig:snap_head_on}
\end{figure*}

\begin{table*}

\caption{Summary of the $N$-body simulations. All runs mentioned are done 4 times and average results are used. (1) GMC mass in units of $\Mc=10^3\,\msun$; (2) ratio of the GMC radius and cluster radius ($\chi=\an/\ac$, where \ac=1.18 pc); (3) impact parameter in units of GMC radius; (4) impact velocity in units of internal cluster velocity dispersion ($\vrms$); (5) and (6) are the encounter distance and velocity,  following from columns (3),(4) and Eq.~\ref{eq:pb} and Eq.~\ref{eq:v}; (7) Figure numbers where the results of the runs are shown.}

\begin{center}
\begin{tabular}{rrrrrrrr}
\hline 

  & (1) & (2) & (3) & (4) & (5) & (6) & (7) \\

      &$\Mn/\Mcl$& $\chi$& $b/\an$& $v_0/\vrms$&$p/\an$         & $\vmax/\vrms$  & Figure \\ \hline 
      &  10      &  1    &  0     & 17.9       &0               & 20             & \ref{fig:head_on}\\
      &          &  2    &  0     & 19.0       &0               & 20             &\\
      &          &  3    &  0     & 19.3       &0               & 20             &\\
      &          &  4    &  0     & 19.5       &0               & 20             &\\
      &          &  5    &  0     & 19.6       &0               & 20             &\\
      &          &  6    &  0     & 19.7       &0               & 20             &\\
      &          &  7    &  0     & 19.7       &0               & 20             &\\
      &          &  8    &  0     & 19.8       &0               & 20             &\\
      &          &  9    &  0     & 19.8       &0               & 20             &\\
      &          & 10    &  0     & 19.8       &0               & 20             &\\
\hline
      & 200     & 1      &   10.8     &      27.7     &     10     &     30  &\ref{fig:de_example} \\
      &         &        &   15.8     &      28.5     &     15     &     30  &\\
      &         &        &   20.8     &      28.9     &     20     &     30  &\\
\hline
      &  10     & 5.8    &       0     &      19.6     &      0     &     20  &\ref{fig:snap_head_on} \\
      & 100     & 5.8    &     3.3     &       14.2    &      3     &     15  &\ref{fig:snap_distant}\\
\hline
      &  100     & 5.8   &  0     & 38.44      &     0           &     40        & \ref{fig:de_p},\ref{fig:de_v},\ref{fig:dedm},\ref{fig:dde}\\
      &          &       &  0     & 16.67      &     0           &     20        & \\
      &          &       &  0     &  4.67      &     0           &     10        & \\
      &          &       &  0.324 & 38.52      &     0.3125      &     40        & \\
      &          &       &  0.371 & 16.85      &     0.3125      &     20        & \\
      &          &       &  0.708 &  5.30      &     0.3125      &     10        & \\
      &          &       &  0.645 & 38.74      &     0.625       &     40        & \\
      &          &       &  0.720 & 17.35      &     0.625       &     20        & \\
      &          &       &  1.117 &  6.72      &     0.625       &     10        & \\
      &          &       &  1.278 & 39.14      &     1.25        &     40        & \\
      &          &       &  1.372 & 18.22      &     1.25        &     20        & \\
      &          &       &  1.722 &  8.71      &     1.25        &     10        & \\
      &          &       &  2.530 & 39.52      &     2.5         &     40        & \\
      &          &       &  2.628 & 19.03      &     2.5         &     20        & \\
      &          &       &  2.914 &  9.09      &     2.5         &     10        & \\
      &          &       &  5.030 & 39.76      &     5           &     40        & \\
      &          &       &  5.122 & 19.53      &     5           &     20        & \\
      &          &       &  5.361 &  9.19      &     5           &     10        & \\
      &          &       & 10.026 & 39.89      &     10          &     40        & \\
      &          &       & 10.107 & 19.79      &     10          &     20        & \\
      &          &       & 10.306 &  9.64      &     10          &     10        & \\

\hline
\end{tabular}

\end{center}
\label{tab:runs}                                                
\end{table*}

\subsection{Description of the runs}
We want to avoid sudden jumps in the contribution of the GMC to the
cluster potential energy at the beginning and at the end of the
simulations. In principle, our GMC should start at infinity, have an
encounter with the star cluster and then continue until
infinity. Since we do not want to evolve the cluster to long before
the encounter, to restrict simulation time and to avoid too much
internal dynamical evolution, we put the GMC initially at a distance
of $D=500\,\ac$ from the cluster and give it zero mass. While the GMC
travels from $D$ to $0.9\,D$, approaching the cluster, we increase its
mass gradually from 0 to $\Mn$. Similarly, after the encounter and
while the GMC travels from $0.9\,D$ to $D$ we decrease its mass
gradually from $\Mn$ to 0. To give the cluster time to get back to
equilibrium and to give stars with velocities higher than the escape
velocity time to leave the cluster we evolve it further for a time
equal to half that elapsed after the encounter.

To test the different analytical expressions of
\S~\ref{sec:analytical} for the energy
gain of the cluster, we perform series of head-on ($p=0$), distant ($p>\an$) and close
encounters with different GMC radii and encounter parameters. The
initial velocity ($v_0$) and impact parameter ($b$) of the GMC are
calculated from the desired values of $p$ and $\vmax$, using the
expressions for gravitational focusing of
\S~\ref{subsec:relations}. The cluster properties are described in
\S~\ref{subsec:cluster}. A summary of all
runs is given in Table~\ref{tab:runs}. The results will be discussed
in \S~\ref{subsec:headon_nbody}-\ref{subsec:dedm}.

All clusters are scaled to $N$-body units, such that $G = M = 1$ and
$\ec = -0.25$, following \cite{1986LNP...267..233H}. Then $\rv$ is the
unit of length and follows from the scaling of the energy, since
$\rv\equiv G\Mcl/(4|E_0|)=1$, where $E_0$ is the total (potential and
kinetic) initial energy of the cluster.  We assume virial equilibrium at the
start of the simulation, which implies $\rv(t=0)=1$.

%_________________________________________________________________
\subsection{Head-on encounters}
\label{subsec:headon_nbody}

We perform $N$-body simulations of head-on encounters between a
cluster and a GMC, both described by a Plummer model. GMCs with
different Plummer radii $\an$ were used to compare simulations to the
analytical result of Eq.~\ref{eq:deheadon}. The energy gain of the
cluster\footnote{We start the cluster in the origin, resulting in a
centre of mass velocity after the passage of the GMC. In the
comparison with the analytical models, when we refer to cluster
energy, we refer to the sum of the {\it internal} potential and {\it
internal} kinetic energy, i.e. the energy after the centre of mass
motion has been subtracted.} is always expressed in units of the
initial cluster energy, i.e. the fractional energy gain ($\de$). Since
we are here interested in the effect of the GMC radius on \de, we fix
the mass of the GMC at a constant value. The results can then be
compared with the predicted value from Eq.~\ref{eq:deheadon}. More
specific, when we divide the energy gain for the different cases by
the value that was predicted by BT98 (e.g. Eq.~\ref{eq:debt}), we
should get the factor $C(\chi)$ of Eq.~\ref{eq:cx}. In
\S~\ref{subsec:dedm} we show that there is no good agreement between
the simulations and the analytical predictions when $\de>>1$. To keep
$\de<1$ for all GMC radii, we choose $\Mn=10^4\,\msun$, which is
somewhat lower than a realistic value for \Mn. We choose integer
values for $\chi$ from 1 to 10, where Eq.~\ref{eq:debt} and
Eq.~\ref{eq:deheadon} should be valid for $\chi=1$ and $\chi\ne1$,
respectively. The encounter velocity for all simulations is $\vmax =
20\,\vrms$, where $\vrms=1\,\kms$ is the internal velocity dispersion
of stars in the cluster.

\begin{figure}
    \includegraphics[width=8cm]{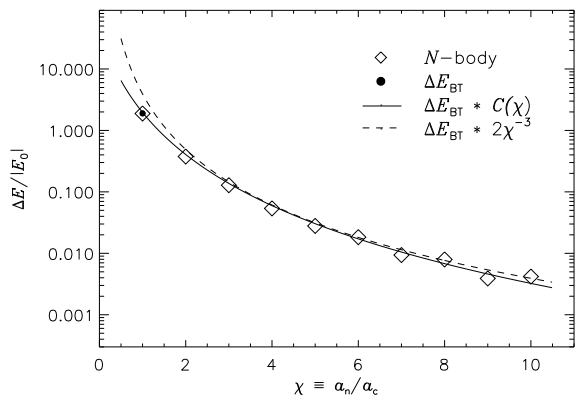} 

    \caption{Fractional energy gain (\de) for a star cluster due to 
    head-on encounters with GMCs with $\Mn=10^4\,\msun$ for
    different $\chi\equiv\an/\ac$ and with $\vmax=20\,\vrms$. The analytical approximation of
    BT87 (Eq.~\ref{eq:debt}) for head-on encounters between identical
    Plummer models is indicated with a filled circle (left). The $N$-body
    results are shown with diamonds. The analytical prediction of
    Eq.~\ref{eq:deheadon} with the correction factor $C(\chi)$ for
    different cloud radii is shown with the full line. The predicted
    energy gain with an approximation for $C(\chi)$ (i.e. $C(\chi)=2\,\chi^{-3}$)
    is shown with the dashed line.}

    \label{fig:head_on}
\end{figure}

In Fig.~\ref{fig:head_on} we show the results of \de\ from the
$N$-body simulations for different $\chi$ with diamond symbols, the
value of BT87 with a filled circle (left) and the result of
Eq.~\ref{eq:deheadon} with the full line. The latter relation is also
shown with a dashed line, but with an approximation for $C(\chi)$ of
the form $C(\chi)=2\,\chi^{-3}$. This approximation is very close to
the real $C(\chi)$ for $2\le\chi\le9$. Excellent agreement between the
analytical work and the $N$-body results is found in all cases.

In Fig.~\ref{fig:snap_head_on} we show snapshots of a cluster during a
head-on encounter with a GMC. In the top panels the line of sight is
perpendicular to the trajectory of the GMC, whose direction is
parallel to the $x$-axis. The attractive force of the GMC parallel to
its trajectory before and after the encounter cancel out, resulting in
displacement of stars only perpendicular to the line of motion of the
GMC. In the bottom panels the line of sight is along the motion of the
GMC, and it can be seen that the stars escape and form a disk like
structure of unbound particles around the cluster. The time with
respect to the encounter moment is indicated in each panel in the top
row, in units of $\tenc$, where $\tenc\equiv2\an/\vmax$.

%_________________________________________________________________
\subsection{Distant and close encounters}
\label{subsec:allp}
\begin{figure}
    \includegraphics[width=8cm]{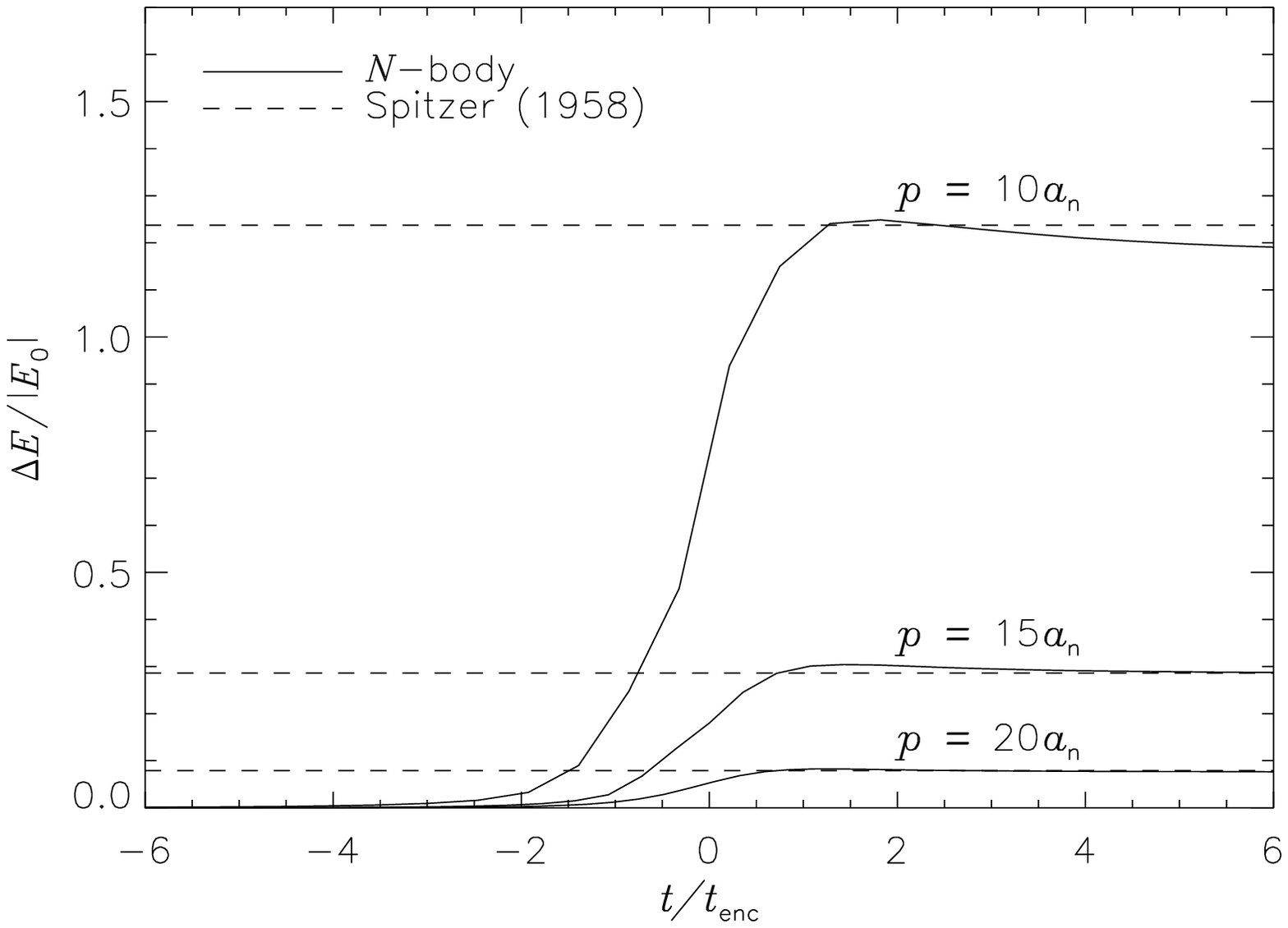} 

    \caption{The evolution of the energy gain of a cluster during
    distant cloud encounters (full lines) with $\Mn=2\times10^5\,\msun$,
    $\vmax=30\,\vrms$ and three different impact parameter $p$. The
    dashed lines indicate the final energy gains predicted by S58
    (Eq.~\ref{eq:despitzer}).}

    \label{fig:de_example}
\end{figure}

\begin{figure*}
    \includegraphics[width=17cm]{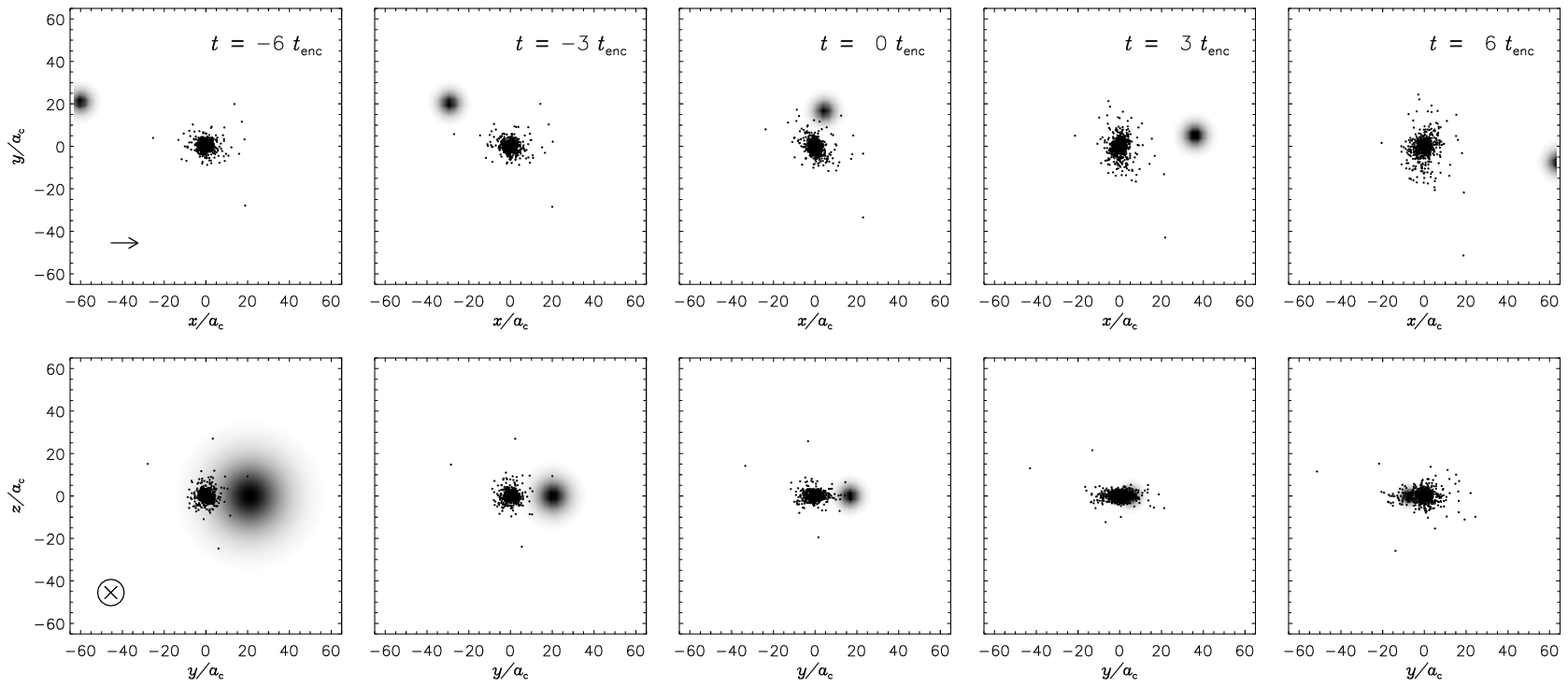}

    \caption{Snapshot of a star cluster during a distant encounter
    with a GMC with $\Mn=100\,\Mcl$, $p=3\,\an$ and $\vmax=15\,\vrms$
    in the centre of mass frame of the cluster. The motion of the GMC
    is along the $x$-axis, starting at $b=y=20\,\ac$, and the line of
    sight is perpendicular to ({\it top panels}) and along ({\it
    bottom panels}) the GMC trajectory.  The arrows in the left lower
    corner of the left panels are parallel to the direction of motion
    of the GMC. The GMC is
    shown with grey shades based on the surface density of a GMC with
    $\an=5.8\,\ac$.  The time with respect to the moment of encounter
    is indicated in each panel of the top row, where $\tenc\equiv
    p/\vmax$. The snapshots are taken in the centre-of-mass frame of
    the cluster and the viewing distance is $60\,\ac$.}

\label{fig:snap_distant}

\end{figure*}

For the simulations of distant encounters we choose $\Mn=10^5\,\msun$.
The tidal forces of the GMC deform the cluster into a cigar like
shape, as can be seen in Fig.~\ref{fig:snap_distant}.  In
Fig.~\ref{fig:de_example} we show with the full lines the evolution of
the internal energy of the cluster for encounters with three different
values for $p$ as a function of time, in units of the encounter time
($\tenc\equiv p/\vmax$). (Note that for head-on encounters
$\tenc\equiv2\an/\vmax$.) We define time ($t$) such that the
moment of closest approach is at $t=0$. The value of
\vmax\ for these runs is $40\,\vrms$ and the GMC radius is equal to the
cluster radius. The GMC radius is chosen small in this example in
order not to have too large values of $p$ when the ratio $p/\an$ is
large. Otherwise, the energy gain would be too small.  The
results of Eq.~\ref{eq:despitzer} for these encounters are overplotted
as the dashed lines. The value of \rrms\ changes during the
simulations due to internal dynamical effects and because of
deformation of the cluster by the GMC. This last effect is clearly
visible in Figs.~\ref{fig:snap_head_on} and
\ref{fig:snap_distant}. Since the result of Eq.~\ref{eq:despitzer} is
very sensitive to the value of
\rrms, we performed tests to find at which time \rrms\ should be taken
to get the best agreement between Eq.~\ref{eq:despitzer} and the
	    simulations.  If we take this value too early,
	    Eq.~\ref{eq:despitzer} will underestimate the energy gain
	    compared to the result of the simulation.  This
	    because dynamical evolution will increase \rrms. When
	    using a value
\rrms\ which is determined during the encounter (i.e. $t\simeq0$), the
energy gain from Eq.~\ref{eq:despitzer} overestimates the true
value. We found that best agreement was found when using \rrms\ at
$t=-2\,\tenc$. From Fig.~\ref{fig:de_example} it can be seen that this
is moment where the energy of the cluster starts to increase, whereas
the cluster energy is almost constant before that moment.

The variations of the relative velocity between the cluster and the GMC
are smaller than the variations in \rrms, and for that reason the
result of Eq.~\ref{eq:despitzer} is less sensitive to the time that we
choose the velocity.  Since most of the energy is injected between
$-\tenc$ and $+\tenc$ (Fig.~\ref{fig:de_example}), the best agreement
is obtained when the relative velocity at $t=0$, e.g. $\vmax$, is used.

\begin{figure}
    \includegraphics[width=8cm]{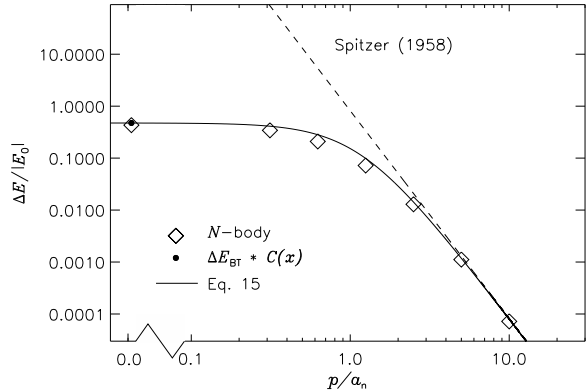}

    \caption{Fractional energy gain of a cluster as a function of the
    impact parameter $p$ and for $\vmax=40\,\vrms$. The results of the
    $N$-body simulations are shown with diamonds. The analytical
    result for head-on encounters, including the correction for the
    radius of the GMC is shown with a filled circle (left). The result
    for distant encounters from S58 is shown with the dashed line.
    The result from the interpolation between the two analytical
    predictions (e.g. Eq.~\ref{eq:despitzer} and \ref{eq:deheadon}),
    i.e. Eq.~\ref{eq:detot}, is shown with the full line.}

    \label{fig:de_p}
\end{figure}

% validation of the interpolation
To test the validity of the interpolation between the distant and the
head-on encounters (Eq.~\ref{eq:detot}), we performed $N$-body
simulations of various encounters with impact parameter $p$ ranging
from 0 to 10 times the GMC radius \an. The encounter velocity was
chosen $40\vrms$, such that the fractional energy gain is always lower
than 1 (Eq.~\ref{eq:deheadon}). The impact parameter $b$ and initial
velocity $v_0$ of the GMC where calculated using Eq.~\ref{eq:pb}. The
GMC mass is $10^5\,\msun$ and we used the observed mass-radius
relation for GMCs (Eq.\ref{eq:mrclouds}), resulting in $\chi=5.8$ and
therefore $C(\chi)=0.01$ (Eq.~\ref{eq:cx}). In Fig.~\ref{fig:de_p} we
show the result of \de\ from the $N$-body simulations (diamonds) and
the analytical prediction of Eq.~\ref{eq:detot} (full line). The
predictions for the two extreme values of $p$, i.e. for distant and
for head-on encounters are shown with the dashed line and the filled
circle (left), respectively. The analytical expression
Eq.~\ref{eq:detot} agrees very well with the results from the $N$-body
simulations. It can be seen that the prediction of S58 is accurate for
$p\ge5\,\an$, in agreement with what was found by
\citet{1985ApJ...295..374A}.  Note that in Fig.~\ref{fig:de_p} we
have used \an\ as a scale length to normalise $p$. This is possible
since GMCs have finite sizes. If we were dealing with objects of
infinitesiaml size, as e.g. black holes, this scaling would not be
possible. Thus, Fig.~\ref{fig:de_p} can not be applied to those
cases. For large $p$ ($p>>\an$) the energy gain only depends on
$p$. This is demonstrated by the good agreement with the point source
approximation of S58, since for these cases $\an=0$.

\begin{figure}
    \includegraphics[width=8cm]{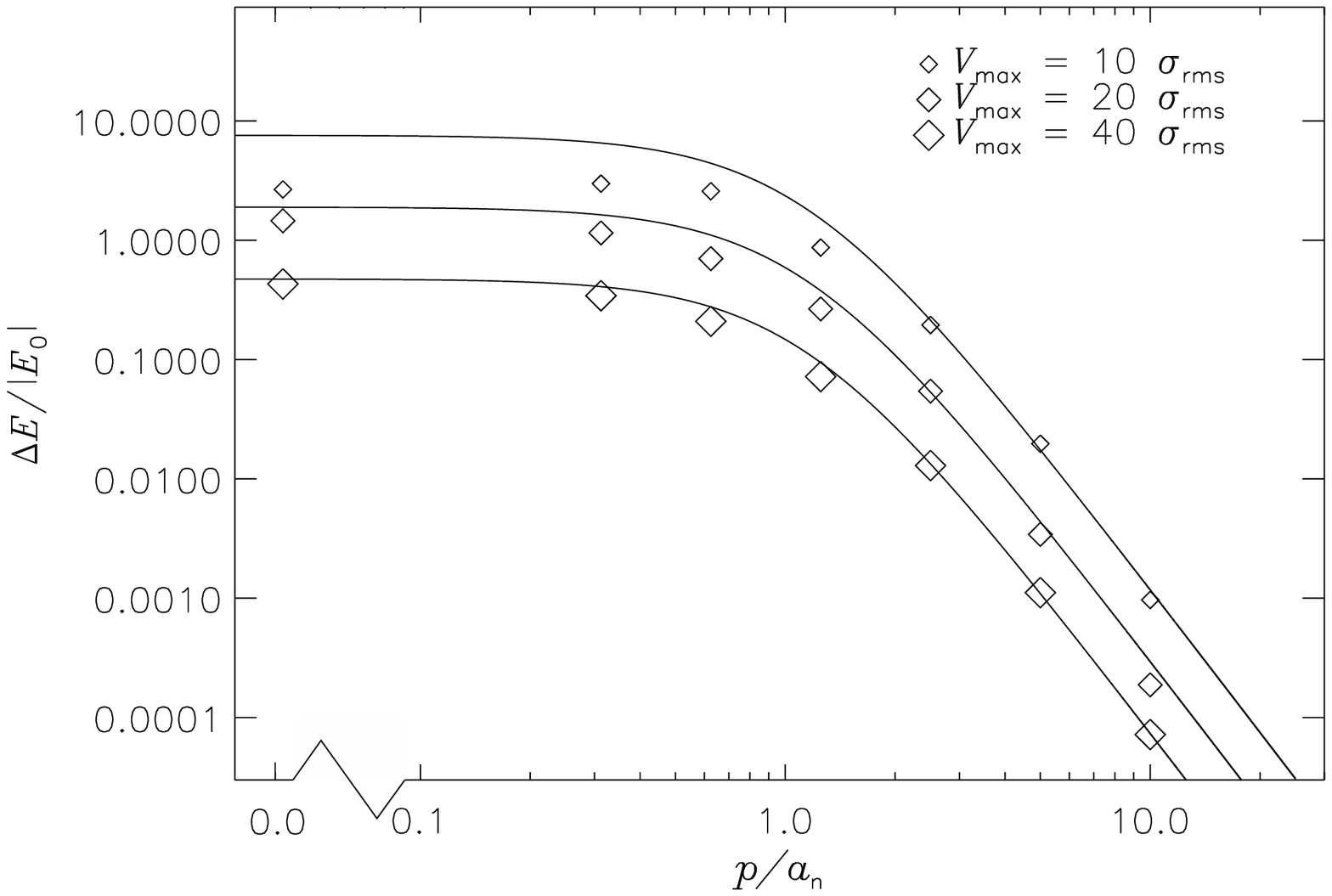} 

    \caption{Fractional energy gain of a cluster as a function of the
    impact parameter $p$ in units of \an\ and for different values of 
    $\vmax$. The results of the $N$-body simulations are shown with
    diamonds, where the size of the diamonds corresponds to the encounter
    velocity. The analytical results from Eq.~\ref{eq:detot} are shown
    with the full lines.}

    \label{fig:de_v} 
\end{figure}

To see if the agreement between our predicted energy gain from
Eq.~\ref{eq:detot} and $N$-body simulations holds for lower
velocities, we performed a similar series $N$-body simulations as
shown in Fig.~\ref{fig:de_p} for $\vmax = 10\vrms$ and $20\vrms$. The
values of \de\ for these three different values of \vmax\ and different values
of $p$ is shown in Fig.~\ref{fig:de_v}. For the lower velocities we
again find very good agreement. However, when the fractional energy gain is
larger than 1, the $N$-body simulations show a smaller energy gain
than the predictions. The difference is larger for larger
energy gains. Similar results were obtained by
\citet{gieles06a} for one-dimensional shocks due to spiral arms. This
is probably because the impulsive approximation does not hold
anymore. When the GMC is very massive, or when the relative velocity
is low, the stars in the cluster are displaced too much before the
encounter. The symmetry assumption of the encounter of the impulse
approximation then does not hold anymore.

The energy gain from the encounter is absorbed mainly by the stars far
from the cluster centre, which can gain velocities much higher than
the escape velocity. A bound core will remain, even when $\de>>1$. Both topics will be discussed in more detail in
\S~\ref{subsec:dedm}.

%_________________________________________________________________
\subsection{Relation between predicted energy gain and mass loss}
\label{subsec:dedm}

\begin{figure}
    \includegraphics[width=8.5cm]{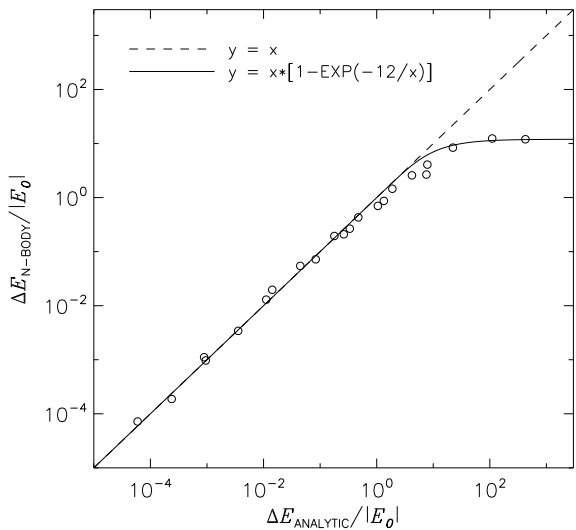} 

    \caption{Relation between the fractional energy gain from the
    simulations and the analytical prediction of Eq.~\ref{eq:detot}
    (circles). A one-to-one relation is overplotted as the dashed
    line. A simple functional form, describing a linear function that
    saturates at a value of 12, fits the result at the high energy
    part well.}

    \label{fig:dde}
\end{figure}

In the previous section we found good agreement between the $N$-body
simulations and our analytical formula for \de\ of a
cluster due to encounters with GMCs with different $p$ and $\vmax$ as
long as $\de<1$ (see Eq.~\ref{eq:detot}). In Fig.~\ref{fig:de_v} we
show that deviations occur whenever the energy gain is larger than the
initial cluster energy. We did a few more simulations of head-on
encounters with higher GMC masses and compare all the predicted energy
gains to the results from the simulations.  The results are shown in
Fig.~\ref{fig:dde}. A one-to-one relation (dashed line) holds until
$\Delta E/|E_0|\simeq5$. A simple functional 
($y=x\,[1-\exp(-12/x)]$), which starts to deviate from a one to one
relation around $y\simeq5$ and levels off at $y=12$ is
also shown in Fig.~\ref{fig:dde} and describes the results well.

\begin{figure}
    \includegraphics[width=8.5cm]{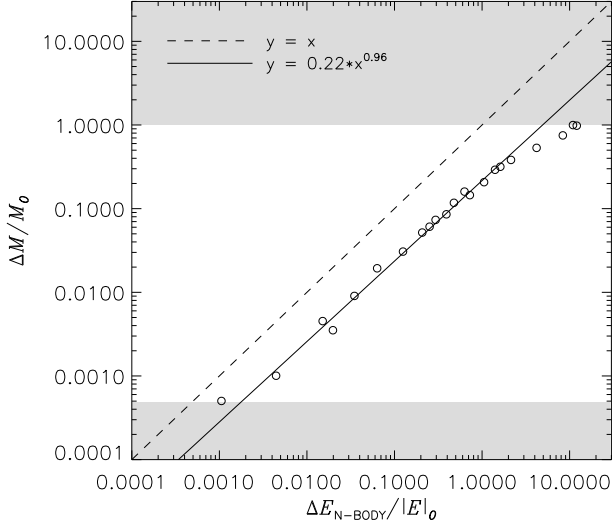} 

    \caption{Relation between the fractional mass loss ($\Delta
    M/M_0$) and the fractional energy gain ($\Delta E/|E_0|$) for the
    simulations of \S~\ref{sec:nbody} (circles). The dashed line shows
    a one-to-one relation and the full line is a fit to the
    simulations. The grey areas indicate the regime where $\Delta
    M/M_0$ can not be defined. }

    \label{fig:dedm}
\end{figure}

To derive a realistic disruption time of clusters due to encounters
with GMCs, we need to know how the energy gain relates to the number
of stars, or mass, that is lost from the cluster in the different
encounters. In Fig.~\ref{fig:dedm} we show the fractional mass loss
($\Delta M/M_0$) as a function of $\Delta E/|E_0|$. The dashed line
indicates a one-to-one relation and the full line is a fit to the
results. The grey areas indicate the regions where $\Delta M/M_0$
cannot be defined, since the cluster can not loose less then one star
(i.e. $M_0/2048$, lower region), and it can not loose more than the
total initial cluster mass (i.e. $M_0$, upper region). The fit shows
an almost linear relation, where \dm\ is a fraction $f=\f$ of
\de. Given the good agreement between the predicted energy gain and
the one from the simulations, we can conclude that a fraction $f=\f$
of the input energy is used to unbind stars. This implies that a
fraction $(1-f)$ of \de\ went into the velocity of the escaping stars,
so that these have escaped with velocities much higher than the local
escape velocity.  Combined with the saturation of
Fig.~\ref{fig:dde}, this shows that some encounters with $\Delta
E\gtrsim10\,E_0$, called overkill encounters
\citep{1985IAUS..113..449W,1988IAUS..126..393W}, can completely
disrupt a cluster. We will further discuss this topic in
\S~\ref{subsec:tdissingle}.

Since we consider isolated clusters, one might argue that there can be
stars that have remained bound, but that would have been pushed over
the tidal radius if the cluster was located in a galaxy. Assuming a
disk galaxy with a flat rotation curve, with rotational velocity of
$V_0 = 220\,\kms$, the tidal radius of a cluster depends then on its
mass and galactocentric distance ($R$) as
$\rt=\left(G\Mcl/2V_0^2\right)^{1/3}\,R^{2/3}$. The estimated values
of $\rt$ for our cluster at $R=[0.5,1,2]\times R_0$, where
$R_0=8.5\,$kpc, i.e. the solar neighbourhood, are $\rt\simeq[13.8,
21.8, 34.6]\,$pc, respectively. We counted for our simulations the
number of extra stars ($\dn$) which after the encounter beyond these
three radii. There is a linear relation between \dn\ and \dm, where
the ratio of the two quantities is $[1.3,1.1,1.0]$ for the three
different tidal limits, respectively. So the effective value of $f$ is
somewhat higher when we take into account the tidal limit of the
cluster. For the solar neighbourhood a good estimate would be
$f=0.22\times1.1\simeq0.25$. For clusters in the centres of galaxies
the tidal radius is small and $f$ can be a factor $\sim1.5$ higher.
From now on we will assume a linear relation $\de=f\,\dm$, with
$\simeq0.25$, which takes into account stars that have become unbound
in isolation and/or are pushed over the tidal limit. This will be used
in
\S~\ref{sec:disruption} to derive the mass loss of clusters in
environment with GMCs and the resulting cluster disruption time.

Hitherto, the cluster disruption time has always been defined as the
time needed to bring the cluster energy to zero by periodically
injecting energy with shocks (e.g. \citet{1972ApJ...176L..51O} for the
disruption time by disk shocks and \citet{1973ApJ...183..565S} for the
disruption by GMCs). We have shown that this overestimates the
disruption time by a factor of $1/f\simeq4$. A more accurate disruption
time, taking into account this factor $f$ is derived in the following
section.

%%%%%%%%%%%%%%%%%%%%%%%%%%%%%%%%%%%%%%%%%%%%%%%%%%%%%%%%%%%%%%%%%%%
\section{Cluster disruption time due to encounters with GMCs}
\label{sec:disruption}

With the description of the mass loss from a cluster due to different
types of encounters with GMCs of
\S\S~\ref{sec:analytical},\ref{sec:nbody} we can derive the disruption
time of star clusters in an environment with GMCs, such as the solar
neighbourhood.

\subsection{Clusters in an environment with GMCs}
\label{subsec:environment}
When considering a distribution of GMCs and clusters with a relative velocity
dispersion $\vcn=\sqrt{\sigma_{\rm n}^2+\sigma_{\rm c}^2}$, the probability of a  relative
velocity in the range $v_0, v_0+\dr v_0$ can be expressed as (BT87, Eq. 7-60)

\begin{equation}
\dr P = \frac{4\pi v_0^2}{(4\pi\vcn^2)^{3/2}}\,\exp{\left(-\frac{v_0^2}{4\vcn^2}\right)}\,\dr v_0.
\end{equation}
Assuming a spatially homogeneous distribution of GMCs, the rate of encounters in the range $b, b+\dr b$ is then

\begin{eqnarray}
\dot{C} & = & \ncloud\,v_0\,2\pi \,b\,\dr b\,\dr P \\
        & = & \frac{\ncloud\,8\pi^2\,b\,\dr b}{(4\pi\vcn^2)^{3/2}}\,\exp{\left(-\frac{v_0^2}{4\vcn^2}\right)}\,v_0^3\,\dr v_0,
\label{eq:encounterrate}
\end{eqnarray}
where $\ncloud$ is the number density of clouds. 
The energy of the cluster increases at a rate 
\begin{equation}
\dot{E}=\int_0^{\infty}\int_{b_{\rm min}}^{b_{\rm max}}\dot{C}(b,v_0)\,\Delta E(b,v_0)\dr b\dr v_0,
%\dot{E}=\int_0^{\infty}\int_{b_{\rm min}}^{b_{\rm max}}\dot{C}\,\Delta E\dr b\,\dr v_0,
\label{eq:energyrate}
\end{equation}
where $\Delta E(b,v_0)$ follows from Eq.~\ref{eq:detot} after
converting $b$ and $v_0$ to $p$ and $\vmax$ due to gravitational
focusing. This integral can be calculated numerically.  

\begin{figure*}
    \includegraphics[width=18cm]{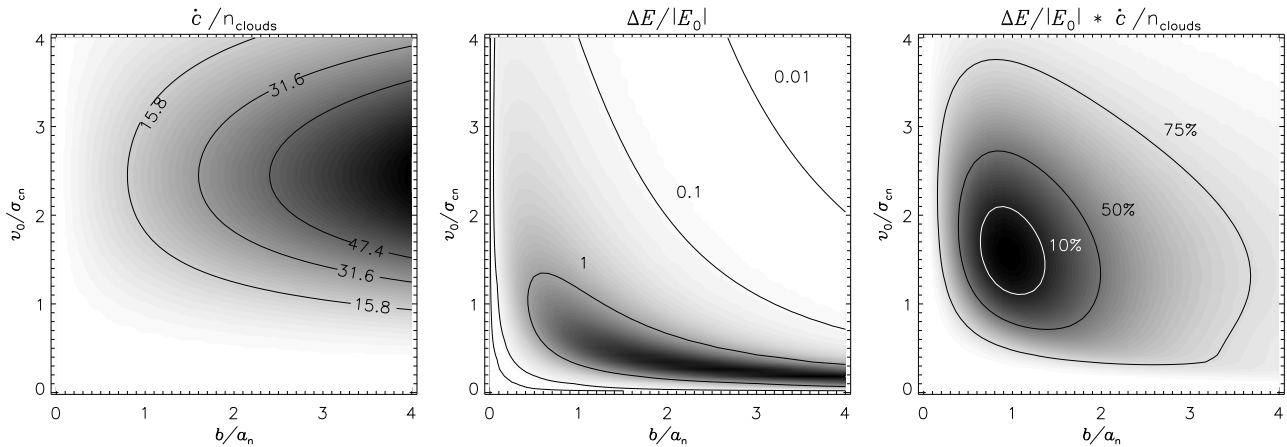} 

    \caption{{\it Left:} The encounter rate ($\dot{C}$,
    Eq.~\ref{eq:encounterrate}), relative to $\ncloud$, for different
    values of $b$ and $v_0$. {\it Middle:} Fractional energy gain
    (Eq.~\ref{eq:detot}) for encounters with different $b$ and
    $v_0$. {\it Right:} Rate of energy increase, resulting from the
    multiplication of $\dot{C}$ and $\Delta E$.}

    \label{fig:encounter_rate}
\end{figure*}

In Fig.~\ref{fig:encounter_rate} we illustrate this using the
parameters for clusters and GMCs used in the $N$-body simulations of
\S~\ref{sec:nbody}, e.g. $\Mn=10^5\,\msun$, \an=6.8\,pc and $\sqrt{\rrms}\simeq
1.8\,$pc.  We assume that young clusters and GMCs have the same
velocity dispersion, i.e. $\sigma_{\rm n}=\sigma_{\rm c}=7\,\kms$
\citep{1984ApJ...281..624S,2006A&A...445..545P}. The velocity
dispersion of the motion of GMCs with respect to clusters, is then
$\vcn=\sqrt{\sigma_{\rm n}^2+\sigma_{\rm c}^2}\simeq10\,\kms$. In the
left panel of Fig.~\ref{fig:encounter_rate} we show the impact rate
$\dot{C}$ of Eq.~\ref{eq:encounterrate} in units of $\ncloud$ as a
function of $b$ and $v_0$, in units of $\an$ and $\vcn$,
respectively. In the middle panel we show \de\ for individual
encounters as a function of $b/\an$ and $v_0/\vcn$, where we have
converted $b$ and $v_0$ to $p$ and $v_0$ first, and then substituted
these values in the equation for energy gain (Eq.~\ref{eq:detot}). The
rate of energy gain as a function of $b/\an$ and $v_0/\vcn$ is the
product of the encounter rate times the energy gain per encounter and
is shown in the right panel. The isocontours indicate the regions
where 10, 50 and 75\% of the total energy per unit time is
produced. This total energy was not determined from this graph, but
from a numeric integration of a similar figure, but with an area that
is 100 times bigger. Most of the energy is injected by fast
($v_0/\vcn\simeq2$) and close ($b/\an\simeq1$) encounters. The
encounter velocities ($\vmax$) are even higher than $2\vcn$ due to
gravitational focusing and this means that the impulse approximation
is valid for these encounters. In
\S~\ref{subsec:allp} we found that the distant encounter regime of
S58 holds for $p\ge5\,\an$. The peak in the right panel of
Fig.~\ref{fig:encounter_rate} around $b\simeq\an$ illustrates that
the approximations of S58 for a point source GMC are not applicable
for the encounters that inject the majority of the energy. In that
regime it is important to use the more accurate description for the
energy gain found in \S~\ref{subsec:detotal} (Eq.~\ref{eq:detot}).

\begin{figure}
    \includegraphics[width=8.5cm]{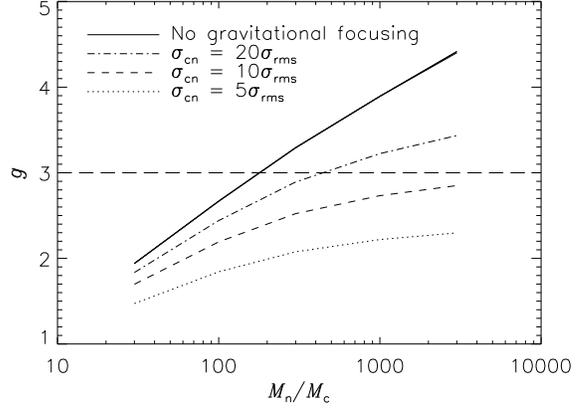} 

    \caption{The correction factor $g$ for close encounters (see text
    for details) as a function of GMC mass. The full line shows the
    result when gravitational focusing is ignored, and the other lines
    show the results for 3 different values of $\vcn$ when
    gravitational focusing is considered. The predicted value of $g=3$ of BT87 is
    indicated with the dashed horizontal line.}

    \label{fig:g}
\end{figure}

BT87 solved Eq.~\ref{eq:energyrate} by substituting the energy gain
for distant encounters of S58. They then integrate $b$ from $\Rn$ to
infinity and correct for the missing close encounters with a factor
$g$ of order unity. The advantage of this solution is that the result
has $\Rn^2$ in the denominator, which can be combined with $\Mn$ and
$\ncloud$ to express the final result in the observable molecular gas
density ($\rhon$) and surface density of individual GMCs
($\sigman$). Our results allow us to quantify the parameter $g$ of
BT87 by integrating Eq.~\ref{eq:energyrate} numerically from 0 to
infinity and from $\Rn$ to infinity. When gravitational focusing is
ignored, the value of $g$ depends only on $\Mn$, since in that case $p=b$
and $\vmax=v_0$. Including gravitational focusing makes $g$ dependent
on $\vcn$ as well, since then encounters with initial $b>\Rn$ can
result in $p<\Rn$. Intuitively one thinks that the inclusion of
gravitational focusing will increase the value of $g$, since there are
more encounters with $p<\Rn$ to correct for. However, the encounter
velocity \vmax\ is always higher than $v_0$, which will make the
energy gain of the cluster for these encounters lower. In
Fig.~\ref{fig:g} we show the resulting $g$ for various values of $\Mn$
ignoring gravitational focusing (full line) and for three values of
$\vcn$. The horizontal dashed line indicates the estimated value of
BT87 for $g$. For the solar neighbourhood (e.g. $\Mn \simeq
10^{5}\,\msun, \vcn
\simeq 10\,\kms$) our results show that $g\simeq 2.5$, slightly lower
than predicted by BT87.

Using the relation between energy gain and mass loss of
\S~\ref{subsec:dedm} we can define the mass loss rate of the cluster as
\begin{equation}
\Mdot=f\frac{\dot{E}}{E}\,\Mc,
\label{eq:rel_dm_de}
\end{equation}
where $f=0.25$ is a dimensionless constant that relates fractional
energy gain to fractional mass loss (see Fig.~\ref{fig:dedm} and
\S~\ref{subsec:dedm}). The initial cluster energy is $E_0 =
-\eta\,GM^2/(2\rh)$, where $\eta\simeq0.4$, depending on the cluster
profile \citep{1987degc.book.....S}.

We can derive a  relation for the mass loss, using
the expression for $\dot{E}$ of BT87 (their Eq.~7-66), where we
substitute $\sigman\equiv\Mn/(\pi\Rn^2)$,  $\rhon\equiv\Mn\,\ncloud$ and $\dot{M}=f\,\dot{E}$
\begin{equation}
\Mdot=\frac{8{\pi}^{3/2}g\,f}{3\vcn\eta}G\left(\sigman\rhon\right)\left(\frac{\rrms}{\rhs}\right)\rh^3,
\label{eq:mdot}
\end{equation}
where the ratio $\left(\rrms/\rhs\right)$ depends on the
cluster model. For the Plummer cluster that we used
$\left(\rrms/\rhs\right)\simeq4$ and for King profiles with
dimensionless potential depth values of $W_0=[3,5,7,9]$ it follows
that $\left(\rrms/\rhs\right)\simeq[1.5,2,3.5,3.7]$.  From
Eq.~\ref{eq:mdot} we can derive the disruption time of clusters, by defining
\begin{equation} 
\tdis = \Mc/\Mdot=\frac{3\vcn\eta}{8{\pi}^{3/2}g\,f\,G}\left(\frac{1}{\sigman\rhon}\right)\left(\frac{\rhs}{\rrms}\right)\frac{\Mc}{\rh^3}.
\label{eq:tdis}
\end{equation} 
For the constants in Eq.~\ref{eq:tdis} we have to make several
assumption. First, we assume $f=0.25$, as was found in
\S~\ref{subsec:dedm}. For $\vcn$ we take $10\,\kms$, based on
observations of GMCs in the solar neighbourhood
\citep{1989ApJ...339..763S} and open clusters
\citep{2006A&A...445..545P}. For the cluster profile we assume a King profile with $W_0 =
3$, corresponding to $c\equiv\log(\rt/\rc)\simeq0.7$, which is the mean
concentration of the galactic open clusters
\citep{1962AJ.....67..471K}. King profiles are more realistic due to
the tidal truncation, which the Plummer model does not have. For the
$W_0=3$ profile considered here, $\left(\rhs/\rrms\right)\simeq0.67$
and $\eta=0.42$ (BT87).

\subsection{Dependence on the relation between cluster mass and radius}
The solution for $\tdis$ of Eq.~\ref{eq:tdis} is very sensitive to the
choice of $\rh$.  Observations of (young) extra-galactic star
clusters show that the projected half-light radius ($\reff$) of the
majority of star clusters is confined to a narrow range of 2-4 pc
(e.g. \citealt{2005A&A...431..905B,2004A&A...416..537L}) and a very
shallow relation between the radius and the mass. A factor of two
difference in radius and nearly equal masses still results in a factor
of eight difference in \tdis. This implies that clusters that are a
few times larger than the mean radius are very unlikely to survive and
cluster that are a few times smaller will be almost insensitive to
encounters by GMCs.  We choose the use the recently observed relation between
radius and mass of
\citet{2004A&A...416..537L}, who fits a function of the form
$\reff=A\,\Mc^{\lambda}$ to a sample of young star clusters in nearby
spiral galaxies and finds that $A=1.12\pm0.35$ and
$\lambda=0.10\pm0.03$. There is a large scatter around the 
fit relation, which, combined with the high sensitivity of \tdis\ on
\rh\ (Eq.~\ref{eq:tdis}), makes it hard to relate mass to
radius. However, since the relation of \citet{2004A&A...416..537L}
reflects the median relation between \reff\ and \Mc, the conversion
will probably hold for a large cluster population.
 To convert $\reff$ to $\rh$ we use the relation $\rh=4/3\,\reff$
\citep{1987degc.book.....S} and with that we can rewrite the result of
\citet{2004A&A...416..537L} as
\begin{equation}
\rh = 3.75\,\left(\frac{\Mc}{10^4\,\msun}\right)^{\lambda},
\label{eq:rm}
\end{equation}
with $\lambda\simeq0.1$. Substituting Eq.~\ref{eq:rm}
in Eq.~\ref{eq:tdis} and inserting all constants, yields an expression for the disruption time
that is independent of $\rh$ and depends on the assumed index that relates
 $\rh$ and $\Mcl$ 

\begin{equation}
\tdis =  2.0\,\left(\frac{5.1\,\msun^2\,{\rm pc^{-5}}}{\sigman\rhon}\right)\left(\frac{\Mc}{10^4\,\msun}\right)^{\gamma}{\mbox{Gyr}},
\label{eq:tdis2}
\end{equation}
with $\gamma=1-3\lambda$. We normalised the product $\sigman\rhon$ to
the value for the solar neighbourhood ($\sigman=170\,\msun$pc$^{-2}$
and $\rhon=0.03\,\msun$pc$^{-3}$, \citealt{1987ApJ...319..730S}). When
taking the 1 sigma upper bound value of $\lambda\simeq0.13$ from
\citet{2004A&A...416..537L}, the index $1-3\lambda=0.61$ agrees very well
with the value of 0.62 found from observations
\citep{2005A&A...429..173L} and from $N$-body simulations of clusters
in the solar neighbourhood, dissolving under the combined effect of a
tidal field, a realistic mass function and stellar evolution
\citep{2003MNRAS.340..227B,2000ApJ...535..759T}.  These
simulations adopted a slightly different relation between \rh\ and \Mc\
since they assumed that at a given galactocentric distance the cluster
size scales with the tidal radius ($\rh
\propto \Mc^{1/3}$).  The multiplicative constant of
Eq.~\ref{eq:tdis2} is a factor of about 7 higher than what was found
from earlier studies (e.g. S58;
\citealt{1973ApJ...183..565S}; BT87). The difference can largely be explained  by our 
factor $f\simeq0.25$ from \S~\ref{subsec:dedm}, that relates
energy gain to mass loss. The remaining difference is
caused by the lower value we found for $g$ and a slightly different
value for the product $\sigman\rhon$.

\subsection{Disruption by single encounters}
\label{subsec:tdissingle}
 In the derivation of \tdis\ in the previous sections we
integrated over the full range of $b$ and $v_0$ in
Eq.~\ref{eq:energyrate} to get an expression for the rate of energy
gain of a cluster due to GMC encounters. Doing so, we implicitely
assume that disruption is caused by a large number of
encounters. However, if the cluster is completely disrupted by the
first encounter with a GMC, Eq.~\ref{eq:tdis} will overestimate
\tdis. Fig.~\ref{fig:dedm} shows that a single encounter can indeed
disrupt the cluster (see also \citealt{1985IAUS..113..449W} and
\citealt{1987MNRAS.224..193T}). If cluster disruption was always caused by
 single encounters, then
Eqs.~\ref{eq:tdis}\&\ref{eq:tdis2} would overestimate the disruption
time in all cases. The most accurate approach is to
include the saturation function of Fig.~\ref{fig:dde} in the relation
between energy gain and mass loss (Eq.~\ref{eq:rel_dm_de}), i.e. to
exclude encounters with $\Delta E \gtrsim 10\,E_0$. This results in \dm=1
for $\de\gtrsim10$. However, this would not have allowed us to come to
the convenient linear dependence of \tdis\ on \rhon\ and \sigmamol\ in
Eq.~\ref{eq:tdis}.

\begin{figure}
    \includegraphics[width=8cm]{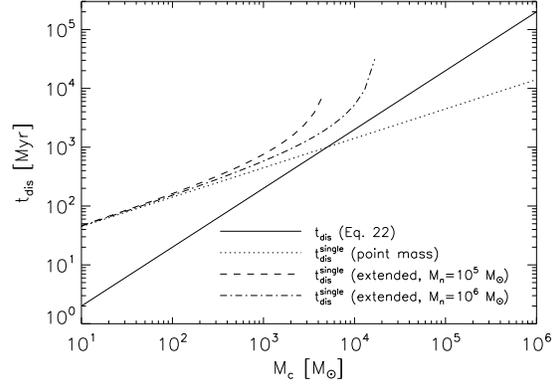}

    \caption{Comparison between \tdis\ of Eq.~\ref{eq:tdis} (full
    line) and the single encounter disruption time
    ($\tdissingle$). The dotted line shows the result of
    \citet{1985IAUS..113..449W} for point-mass GMCs and the dashed and
    dotted-dashed lines show the results for extended GMCs with $\Mn=10^5\,\msun$ and $10^6\,\msun$, respectively.}

    \label{fig:tdissingle}
\end{figure}

Alternatively, we here estimate the typical time scale for a cluster
to be disrupted by one single encounter ($\tdissingle$).  Following
\citet{1985IAUS..113..449W}, we can calculate the critical encounter
distance $(\pcrit)$ for which one encounter will result in total
disruption of the cluster. Based on Eq.~\ref{eq:detotrh}, on the
expression for the cluster energy $|E_0|=\eta\,G\Mc^2/(2\rh)$ and on
the relation between
\de\ and \dm\ from \S~\ref{subsec:dedm} we can derive \pcrit\ by
solving for $p$ in $\de=1/f$, such that
\begin{equation}
\pcrit^2=\left(\frac{8Gf}{3\eta}\frac{\rrms}{\rhs}\right)^{1/2}\,\frac{\Mn}{V}\left(\frac{\Mc}{\rh^3}\right)^{-1/2}-K.
\label{eq:pcrit}
\end{equation}
In here $K=\sqrt{\rh\,\Rn^{3}}$, which is the term in the denominator
of Eq.~\ref{eq:detotrh} that makes $\Delta E$ converge to the correct
value for head-on encounters ($p=0$). \citet{1985IAUS..113..449W} did
not include this term, since he assumed point-mass GMCs
(Eq.~\ref{eq:despitzer}). (Note that Eq.~\ref{eq:detotrh} is equal to
Eq.~\ref{eq:despitzer} when $K=0$). The number of encounters with
encounter parameter $p$ depends on the GMC number density ($\ncloud$)
as

\begin{equation}
N=\pi p^2\,\ncloud\,V\,\Delta t.
\end{equation}
We can solve for $\Delta t$ by setting $N=1$ and $p=\pcrit$, which is
then the time it takes for a disruptive encounter to occur
\begin{equation}
\tdissingle = \frac{1}{\pi\pcrit^2\,\ncloud\,V}.
\label{eq:tdissingle1}
\end{equation}
In here gravitational focusing is ignored. Following
\citet{1985IAUS..113..449W} and assuming point-mass GMCs, e.g. $K=0$,
we find an expression of
\tdissingle\ of the form

\begin{equation}
\tdissingle=\left(\frac{3\eta}{8Gf\pi^2}\frac{\rhs}{\rrms}\right)^{1/2}\left(\frac{\Mc}{\rh^3}\right)^{1/2}\frac{1}{\rhon}
\label{eq:tdissingle},
\end{equation}
where we have used $\rhon\equiv \Mn\ncloud$ again. At first sight this
relation for the disruption time looks similar to what we derived
in the previous section (Eq.~\ref{eq:tdis}). The main difference is
the scaling with the cluster density ($\Mc/\rh^3$). In Fig.~\ref{fig:tdissingle} we
show a comparison between \tdissingle\ of Eq.~\ref{eq:tdissingle}
(dotted line) and \tdis\ of Eq.~\ref{eq:tdis} (full line) for different
\Mc. Here we used a constant \rh=3.75 pc and the same values for $f,
\eta$ and $(\rrms/\rhs)$ as in
\S~\ref{sec:disruption}. The value of \tdissingle\ is smaller than the
result of Eq.~\ref{eq:tdis2} for clusters with
$\Mc>5\times10^3\,\msun$. This implies that for massive clusters the
life time is limited by the first disruptive encounter, rather than
the successive heating by multiple encounters. For cluster of lower
mass, the time it takes to disrupt by encounters with $p>\pcrit$ is
shorter than \tdissingle. This mass dependence is counter intuitive at
first, but is explained by the difference in scaling with $\Mc$
(Eqs.~\ref{eq:tdis}\&\ref{eq:tdissingle}). Note that the crossing
point (Fig.~\ref{fig:tdissingle}) does not depend on \rhon, since
Eqs.~\ref{eq:tdis}\&\ref{eq:tdissingle} depend in the same way on
\rhon. 

In Fig.~\ref{fig:tdissingle} we also plot the result of \tdissingle\
by substituting Eq.~\ref{eq:pcrit} with $K=\sqrt{\rh\Rn^3}$ in
Eq.~\ref{eq:tdissingle1}, based on $\Mn=10^5\,\msun$ (dashed line) and
$\Mn=10^6\,\msun$ (dotted-dashed line) and Eq.~\ref{eq:mrclouds}. Here
we use the same value for \rhon, and then assume that all mass is in
clouds with $\Mn=10^5\,\msun$ and $10^6\,\msun$, respectively, to
solve for $\ncloud$ (Eq.~\ref{eq:tdissingle1}).  For low mass
clusters, these solutions for \tdissingle\ agre with the result of
\citet{1985IAUS..113..449W}, since there $\pcrit\gtrsim {\rm few}\Rn$,
where the assumption of point-mass GMCs is valid (see
Fig~\ref{fig:de_p}). For clusters with $10^2<\Mc/\msun<10^4$, the
solution for \tdissingle\ is higher than that of the point-mass
assumption. This is because \pcrit\ is smaller when extended GMCs are
considered. In close encounters the extent of the GMC, e.g. $K$, is
more important (Eq.~\ref{eq:detot}). There is no solution for clusters
with $\Mc>5-10\times10^4\,\msun$, meaning that it is not possible to
disrupt these clusters by a single encounters with a GMC with masses
up to few times $10^6\,\msun$. 

The fact that the solutions for \tdissingle\ for realistic (extended)
GMCs are always larger than
\tdis\ from Eq.~\ref{eq:tdis} justifies our assumptions of
\S~\ref{subsec:environment}. However, Fig.~\ref{fig:tdissingle} shows
that for a cluster with $\Mc=10^{3-4}\,\msun$ the two solutions are
close, meaning that clusters are destroyed by just a few
encounters. This means that individual clusters will have strongly
varying lifetimes. Our results of \S~\ref{sec:disruption} will,
however, be valid as a statistical mean.

%%%%%%%%%%%%%%%%%%%%%%%%%%%%%%%%%%%%%%%%%%%%%%%%%%%%%%%%%%%%%%%%%%%
\section{Conclusions and discussion}
\label{sec:conclusion}

We have derive an expression for the cluster disruption time ($\tdis$)
due to encounters with GMCs, based on the results of $N$-body
simulations of different types of encounters between a cluster and a
GMC. The expression for \tdis\ (Eq.~\ref{eq:tdis2}) allows us to
estimate the disruptive effect of GMCs in different environments and
to compare it to other effects such as the galactic tidal field, stellar
evolution and spiral arm shocks. From Eq.~\ref{eq:tdis2} we see that
the disruption time for a $10^4\,\msun$ cluster in the solar
neighbourhood is $t_4 = 2.0\,$Gyr. This is
a factor of 3.5 shorter than what was found by
\citet{2003MNRAS.340..227B}  for the combined effect of a
realistic mass function, stellar evolution and the galactic tidal
field ($t_4= 6.9\,$Gyr). Note that the index $\gamma$ is comparable,
if we take the upper bound value of $\lambda=0.10\pm0.03$, since
\citet{2003MNRAS.340..227B} found
$\tdis\propto\left[\Mcl/\ln(\Mcl)\right]^{0.75}$. \citet{2005A&A...429..173L}
showed that $\left[\Mcl/\ln(\Mcl)\right]^{0.75}\propto
\Mcl^{0.62}$. The same mass dependence was also found from the age and
mass distribution of cluster samples in different galaxies by
\citet{2003MNRAS.338..717B} and from the age distribution of clusters
in the solar neighbourhood by \citet{2005A&A...441..117L}. For this
last sample they found a shorter scaling value of $t_4=1.3\,$Gyr. 
Eq.~\ref{eq:tdis2} combined with the observed relation between
$\rh$ and $\Mcl$ (Eq.~\ref{eq:rm}) suggests that the combined mass
loss by GMC encounters and evaporation in a tidal field preserves the
index $\gamma\simeq0.62$. A similar scaling between $\tdis$ and
$\Mcl$ was found for the disruption time by spiral arm passages. The
combination of all these effects and a comparison to the observed
age distribution of open clusters in the solar neighbourhood will be
discussed in more detail by
\citet{lamers06}.

The result of Eq.~\ref{eq:tdis2} can easily be applied to
environments/galaxies with different molecular gas densities
($\sigmamol$). The mid-plane density ($\rhon$) of GMCs scales with
$\sigmamol$ as $\rhon=\sigmamol/2h$, where $h$ is the vertical scale
length of molecular gas, which is around $100\,$pc in the solar
neighbourhood \citep{1987ApJ...322..706D}. Assuming that $h$ is the
same in M51, and therefore $\rhon=\sigmamol/200\,$pc, we can predict
$t_4$ due to GMCs encounters once we know $\sigmamol$ in M51 and
compare this to the value of $t_4$ for clusters in M51
\citep{2005A&A...441..949G}. \citet{1993A&A...274..123G} find
$\sigmamol=90\,\msun\,$pc$^{-2}$ for the central region of M51, which
is 14 times higher than in the solar neighbourhood
\citep{1987ApJ...319..730S}. This implies
$t_4\simeq2100\,{\rm Myr}/14=150\,$Myr. Note that this value is an
upper limit, since we have assumed that the surface density of
individual GMCs ($\sigman$) is independent of $\sigmamol$, while this
could be higher in environments with higher $\sigmamol$. In addition,
the values of $f$ and $g$ in Eq.~\ref{eq:tdis} will probably be higher
in the centre of M51 (see \S~\ref{subsec:dedm} and
\S~\ref{sec:disruption}), which will make \tdis\ shorter.  Also,
we have assumed the same density profile as for the Galactic open
cluster, e.g. $W_0 = 3$. A density profile of $W_0=7$, as found for
the Galactic globular clusters, would reduce \tdis\  by a
factor of two (Eq.~\ref{sec:disruption}).  The resulting value of
$t_4$, however, is very close to the value of $t_4=100-200\,$Myr,
determined observationally by
\citet{2005A&A...441..949G} from the age and mass distribution of star
clusters in M51. Based on this work we think that we can explain the
short disruption times of clusters in the centre of M51, if GMC
encounters are the dominant disruption effect.

%A last (extreme) example is the merging galaxy Arp 220, for which
%\citet*{1997ApJ...484..702S} find
%$\sigmamol\simeq2-5\times10^4\,\msun\,$pc$^{-2}$. Since many
%assumptions we have made before do not hold in the case of merging
%galaxies we can not simply plug this value into
%Eq.~\ref{eq:tdis2}. The stunning value for $\sigmamol$ still suggests
%that low mass clusters ($\Mcl \simeq 10^{3-4}\,\msun$) are not likely
%to survive longer than a few tens of Myr. Since the merging phase of
%Arp 220 is short lasting ($t_{\rm merge}$ = few 100 Myr), it implies
%that clusters with masses $\Mcl<10^4\,\left(t_{\rm
%merge}/t_4\right)^{1/\gamma}\,\msun$ will not survive the merging
%phase.

\section*{Acknowledgement}
We thank the referee, Roland Wielen, for useful comments and pointing
us to the problem of single disruptive encounters
(\S~\ref{subsec:tdissingle}). We enjoyed discussions with Nate
Bastian, S{\o}ren Larsen and Sverre Aarseth. This work was done with
financial support of the Royal Netherlands Academy of Arts and
Sciences (KNAW), the Dutch Research School for Astrophysics (NOVA,
grant 10.10.1.11 to HJGLM Lamers) and the Netherlands Organisation for
Scientific Research (NWO, grant 635.000.001).  The simulations were
done on the MoDeStA platform in Amsterdam. MG thanks the Sterrekundig
Instituut ``Anton Pannenkoek'' of the University of Amsterdam for
their hospitality during many pleasant visits.

\bibliographystyle{mn2e}

\end{document}